\documentclass[12pt]{article}
\usepackage[colorlinks]{hyperref}
\usepackage{amsmath}
\usepackage{amsfonts}
\usepackage{graphicx}
\usepackage[latin1]{inputenc}
\begin{document}
\hypersetup{
colorlinks=true,
urlcolor=blue,
linkcolor=blue,
citecolor=blue,
}
\newcommand \fr{``flux rule''}
\begin{center}
{\Large\bf Electromagnetic induction:\\ how the \fr has superseded Maxwell's general law}\\
\vskip2mm
{Giuseppe Giuliani}\\
\vskip2mm
{Formerly, Dipartimento di Fisica, Universit\`a di Pavia - Retired}\\
\vskip2mm
{giuseppe.giuliani@unipv.it}
\end{center}

\begin{abstract}
As documented by textbooks, the teaching of electromagnetic
induction in university and high school courses is primarily based
on what Feynman labeled as the ``flux rule'', downgrading it from
the status of physical law. However,  Maxwell derived a ``general
law of electromagnetic induction'' in which the vector potential
plays a fundamental role. A modern reformulation of Maxwell's law
can be easily obtained by defining the induced electromotive force
as $\oint_l (\vec E +\vec v_c\times \vec B)\cdot\vec{dl}$, where
$\vec v_c$ is the velocity of the positive charges which,
by convention, are the current carriers. Maxwell did not possess a
model for the electric current. Therefore, in his law, he took $\vec v_c$ to be the velocity
 of the circuit element containing the charges.
 This paper
aims to show that the modern reformulation of Maxwell's law
governs electromagnetic induction, and that the \fr is not a
physical law but only a calculation shortcut that does not always
yield the correct predictions. The paper also tries to understand
why Maxwell's  law has been ignored, and how the ``flux rule'' has
taken root. Finally, a section is dedicated to
teaching this modern reformulation of Maxwell's law in high
schools and elementary physics courses.

\end{abstract}
\section{Introduction}
Electromagnetic induction was discovered by Michael Faraday in
1831, about ten years after
Christian {\O}rsted showed that electric currents create magnetic effects.
Faraday induced  currents in a closed
conducting loop either by switching on and off the current in a nearby
circuit (volta-electric-induction),   or by
moving the conducting loop towards or away from a magnet
(magneto-electric-induction) \cite{faraday1}.   In textbooks or classrooms, these
experiments are usually the starting point for a discussion about
electromagnetic induction.
\par
Many physicists tackled the problem experimentally and
theoretically in the years following Faraday's discovery. On the
experimental side, it was not so easy to add novel knowledge to
what Faraday had already discovered. An exception was the rule
found by Emil  Lenz in 1834:  the induced current opposes the
phenomenon that generated it \cite{lenz}.
\par
We must wait for Maxwell's {\em Treatise} in 1873 to find a {\em general}
law of electromagnetic induction, derived within a field
description and by treating the currents with a Lagrangian
formalism  \cite[pp. 207-211]{treatise2}.  The amazing feature of this
law is that it was obtained without knowing what an electric
current is, apart from recognizing that it is a ``kinetic process''.
\par
Realizing that an electric current arises from moving charges,  we can reformulate Maxwell's general law  as an integral around a closed curve $l$:
\begin{equation}\label{generalintro}
 \mathcal E = \oint_l \left(- \frac{\partial \vec A}{\partial t} +
\vec v_c \times \vec B\right)\cdot \vec{dl},
\end{equation}
where $\vec B$ is the external magnetic field to which the circuit is exposed, $\vec A$ is the corresponding vector potential, and $\vec v_c$ is the
velocity of the positive charges that, by convention, are the
current carriers. Maxwell did not possess a model for
the electric current. Therefore, in his law, he took  $\vec v_c$ to be the velocity of the circuit
element containing the charges.
\par
 Maxwell's law fell rapidly into oblivion;
meanwhile, the ``flux rule'' took root. The ``flux rule'' states that the induced
electromotive force ({\em emf}) $\mathcal{E}$ is given by:
\begin{equation}\label{flusso}
\mathcal E=-\frac{d}{dt}\int_S \vec B\cdot \hat n\, dS=-\frac{d\Phi}{dt},
\end{equation}
where $\Phi$ is the magnetic flux, and $\hat{n}$ is the unit vector normal to the integration surface element $dS$. There is no constraint on $S$, except that it is bounded by the conducting circuit.
  Maxwell enunciated this rule about fifty pages before
the formulation of the general law and, regrettably, did not
comment on the relationship between the two \cite[p. 167]{treatise2}. This fact
may have contributed to the oblivion of Maxwell's law (section \ref{why}).
\par
Thereby nowadays, with few exceptions, textbooks for university and high school students consider the \fr to be the law of electromagnetic induction, even though, starting with  Feynman's {\em Lectures}, this view
  has been challenged \cite{feyn2}.
\par
The  present paper aims to prove that the modern formulation of
Maxwell's general law (Eq.\ref{generalintro}) explains all known experiments on
electromagnetic induction. Furthermore, it examines why
the ``flux rule'' has superseded Maxwell's law. It is organized as
follows. Section \ref{maxind} shows how Maxwell derived his law.
Section \ref{mle}  recalls how Maxwell's  law can be reformulated
within Maxwell-Lorentz-Einstein electromagnetism to include our modern knowledge of the nature of electric currents. Section \ref{where} deals with the
localization of the induced {\em emf}. Section \ref{rulesec} shows why
the ``flux rule'' is not a physical law but a calculation shortcut.
Section \ref{why} tries to understand how and why Maxwell's
general law has been neglected and how the \fr has taken root. Finally, in the last section, we discuss the issue of how
to introduce the vector potential and the modern reformulation of
Maxwell's general law in elementary physics courses.
\section{Maxwell and the electromagnetic induction}\label{maxind}
In this section, we shall discuss Maxwell's approach to
electromagnetic induction which he developed in the second volume of his {\em
Treatise}. Unless otherwise stated, the symbols used to denote
physical quantities are the modern ones.  In the introductory and
descriptive part  dedicated to electromagnetic induction \cite[pp.
163-167]{treatise2},  Maxwell enunciated the ``flux rule'', without
formally writing the corresponding equation \cite[p.
167]{treatise2}. Maxwell did not possess a microscopic  model of
the electric current: the corpuscular and discrete nature of the
electric charge was determined in the late nineteenth
century, after the discovery of the electron by Joseph John
Thomson (1897). Hence, Maxwell only observed that ``The electric current
cannot be conceived except as a kinetic phenomenon'' and,
consequently, that ``all that we assume here is that the electric
current involves motion of some kind'' \cite[pp.
196-197]{treatise2}.
\par
The simple idea that  current is a kinetic phenomenon, allowed
Maxwell to treat electric circuits with the Lagrangian formalism
and to use a mechanical analogy. Considering a system of two
filiform (meaning ``threadlike," or one-dimensional) circuits, he wrote that the kinetic energy of the system,
due to the current flowing through  them -- i.e. its {\em
electrokinetic energy} -- is given by \cite[p. 207]{treatise2}:
\begin{equation}\label{cincircuiti}
T=\frac{1}{2}L_1{\dot{y_1}}^2 + \frac{1}{2}L_2{\dot{y_2}}^2 + M_{12}\dot{y_1}\dot{y_2},
\end{equation}
where $\dot{y_1}$ and $\dot{y_2}$ are the currents in the circuits.
In $L_1, L_2$ and $M_{12}$ we recognize, as Maxwell did, the
self-inductances of the two circuits and their mutual inductance.
Then, the {\em electrokinetic momentum} of, for instance, circuit
$\sharp 2$ is given by:
\begin{equation}\label{momele}
p_2=\frac{dT}{d\dot{y_2}}= L_2\dot{y_2}+M_{12}\dot{y_1}.
\end{equation}
Maxwell then first considered the situation where there is only a single conducting filiform loop of resistance $R$
 that, at the instant
$t=0$, is connected to the poles of a battery (whose {\em emf} is $\mathcal E$) and wrote:
\begin{equation}\label{indumax}
\mathcal E= R\dot{y}+ \frac{dp}{dt},
\end{equation}
which, in modern notation, reads (as we teach our students):
\begin{equation}\label{indumax2}
    \mathcal E= Ri+L\frac{di}{dt},
\end{equation}
where $-Ldi/dt$ is the self-induced {\em emf} in the circuit.
\par
 Maxwell's commented \cite[pp. 208-209]{treatise2}:
\begin{quote}\small
    The impressed electromotive force $\mathcal E$ is therefore the sum of two
parts. The first, $R\dot{y}$, is required to maintain the current $\dot{y}$ against
the resistance $R$. The second part is required to increase the electromagnetic momentum $p$.
This is the electromotive force which
must be supplied from sources independent of magneto - electric
induction. The electromotive - force arising from magneto - electric
induction alone is evidently $-{dp}/{dt}$, or, {\em the rate of decrease of the
electrokinetic momentum of the circuit} [original italics].
    \end{quote}
    Since $p=\oint \vec A\cdot\vec {dl}$ \cite[p. 215]{treatise2}, we finally get:
\begin{equation}\label{emfdef}
    \mathcal E= -\frac{dp}{dt}=-\frac{d}{dt}\oint_l \vec A\cdot \vec{dl},
\end{equation}
where $\mathcal E$ has the dimensions of an electric potential, as it should.
\par
Eq.  (\ref{emfdef}) is also valid in the case of two or more circuits,
as it can be  verified. In a section entitled {\em Exploration of the
field by means of the secondary circuit} \cite[p. 212]{treatise2},
Maxwell treated in detail the case of two circuits. Maxwell begun by recalling that, on the basis of Eq.
(\ref{momele}), the electrokinetic momentum of the secondary
circuit consists of two parts. The interaction part is given by:
\begin{equation}\label{momesec}
p_2= Mi_1;\qquad M_{12}=M_{21}=M.
\end{equation}
Maxwell considered only the effect of the circuit $\sharp 1$ on circuit $\sharp 2$ and ignored the self-induction of circuit $\sharp 2$.
Under the assumptions that {\em the primary circuit {\em (circuit
$\sharp1$)} is fixed, and its current $i_1$ is constant},  the electrokinetic
momentum of the secondary circuit depends only -- through the
mutual inductance $M$ --  on its form and position. Hence \cite[p. 216]{treatise2}:
\begin{equation}\label{flusso2}
p_2=\oint_{l_2}    \vec A_1\cdot \vec{dl_2} = \oint_{S_2} \vec B_1 \cdot \hat n\, dS_2,
\end{equation}
where $\vec A_1$ and $\vec B_1$ are the vector potential and the magnetic field created by the first circuit. Now, if the secondary circuit
is at rest, by combining Eq. (\ref{emfdef}) and (\ref{flusso2}), we
get the ``flux rule'' (\ref{flusso}).
\par
 But Maxwell did not make this step.
Instead,  Eq. (\ref{emfdef}) is the starting point to obtain  the {\em General
Equations of the Electromotive Force} \cite[p. 220]{treatise2}. For this, Maxwell considered the secondary circuit to be in motion
with a velocity that can depend on each circuit element, which means that the
circuit may  change form. The {\em emf} induced in the
secondary circuit is given by:
\begin{equation}\label{partenza}
\mathcal E_2= -\frac{dp_2}{dt}=-\frac{d}{dt}\oint_{l_2}\vec A_1(\vec r, t)\cdot \vec{d{l_2}}.
\end{equation}
We have introduced the subscripts for  clarity: Maxwell did not use them.
Maxwell outlined only the main passages of the ensuing
calculation; a recent, detailed derivation can be found in \cite{max_598}. It turns out that (in modern notation):
\begin{equation}\label{leggegenmax}
\mathcal E_2= -\frac{dp_2}{dt}=\oint_{l_2} \left[(\vec v_2 \times \vec B_1) -\frac{\partial \vec A_1}{\partial t}-\nabla\varphi \right]\cdot \vec {dl_2},
\end{equation}
where  $\vec v_2$ is the velocity of the circuit element $dl_2$ and $\varphi$ is the electric potential at the same circuit element. We have left $\varphi$ without subscript because its values at the points of circuit $\sharp 2$ depend both on circuit $\sharp 1$ - through the current induced in circuit $\sharp 2$ - and on the circuit $\sharp 2$ through its electric resistance.
\par
In this calculation, the magnetic field $\vec B_1$ comes into play through the
relationship $\vec B_1=\nabla\times\vec A_1$. The vector:
\begin{equation}\label{electricgen}
\left[(\vec v_2 \times \vec B_1) -\frac{\partial \vec A_1}{\partial t}-\nabla\varphi \right],
\end{equation}
``represents the electromotive force per unit length acting on the
element $dl_2$ of the circuit'' \cite[ p. 222]{treatise2}.  Maxwell
rewrote Eq. (\ref{leggegenmax}) in a form that we translate as:
\begin{equation}\label{after}
\mathcal E_2= \oint \vec E_{eff}\cdot\vec {dl_2},
\end{equation}
where what we have denoted by $\vec E_{eff}$ was represented by
Maxwell with the same symbol $\aleph$ \cite{symbol}
 used for defining the vector
``electric intensity'' \cite[p. 72]{treatise1}.This denotation is inappropriate because
the ``electric intensity'' (electric field) must be a solution of
Maxwell's equations, while $\vec E_{eff}$ is what will  later be
called the Lorentz force on a unit positive charge. The
following comments show that Maxwell implicitly agreed with our
interpretation, despite the use of the symbol
$\aleph$ (in the following quotations we have added $\vec E_{eff}$ next to each symbol for clarity):
\begin{quote}\small
The vector $\aleph$
($\vec E_{eff}$) is the electromotive force at the moving
element $dl_2$.
\par
[\dots]
\par
The electromotive force  at a point has already been defined in \S
68. It is also called the resultant electrical force, {\em being the
force which would be experienced by a unit of positive electricity
placed at that point}.
\par
[\dots]
\par
{\em The electromotive force at a point, or on a particle\em, must
be carefully distinguished from the electromotive force along an
arc of a curve, the latter quantity being the line-integral of the
former}. See \S 69 \cite[pp. 222-223]{treatise2}).
\end{quote}
\par
These  specifications confirm that Maxwell indeed considered the vector of Eq.
(\ref{electricgen}) as the force exerted by the electromagnetic field
on a unit positive charge: it is an extension of the electrostatic
force.  Likely for  this reason, Maxwell denoted the two forces
inappropriately with the same symbol.  \par Going into details,
Maxwell stated that:
\begin{itemize}
\item The first term $\vec v_2\times \vec B_1$ is due to ``the motion
    of the particle through the magnetic field'' \cite[p.
    223]{treatise2}. In this passage, the motion of the ``particle'' is
    identified with the motion of  the ``moving circuit's element''.
    Therefore, in  Eq. (\ref{electricgen}), the velocity $\vec v_2$ is
    attributed to a unit of positive charge. It would
    have required a microscopic model for the electric current to note that the velocity of a charge  is the sum
    of the velocity of the circuit element that contains the charge
    and of the drift velocity of the charge (see section \ref{mle}).
\item The second term $-\partial \vec A_1/\partial t$ ``depends on
    the time variation of the magnetic field. This may be due
    either to the time-variation of the electric current in the
    primary circuit, or to motion of the primary circuit'' \cite[p. 223]{treatise2}.
\item The third term $\nabla\varphi$ is introduced ``for the sake
    of giving generality to the expression for $\vec E_{eff}$.
    It disappears from the integral when extended round the
    closed circuit. The quantity $\varphi$ is therefore
    indeterminate as far as regards the problem now before us, in
    which the total electromotive force round the circuit is to be
    determined. We shall find, however, that when we know all the
    circumstances of the problem, we can assign a definite value
    to $\varphi$, and that it represents, according to a certain
    definition, the electric potential at the point ($x, y, z$)'' \cite[p. 222]{treatise2}.
\end{itemize}
According to Maxwell, the induced electromotive force is
given by the integral over the complete circuit of the force exerted
by the electromagnetic field on a unit positive charge.
 \par
Maxwell's long derivation of the law of electromagnetic induction is very complicated. This feature has probably led Maxwell's contemporaries to overlook it and ignore the physical novelties it contained. Furthermore, the choice of using the Lagrange formalism -- dictated by Maxwell's ignorance of what an electric current is -- likely rendered the derivation too abstract for the tastes of his contemporaries. In the following years, and until today,  textbook writers, with the exception of Bouasse \cite{bouasse},  ignored Maxwell's general law (see section \ref{why} and \cite{SM}).
\par
The following section will straightforwardly re-derive Maxwell's law by correctly defining the induced electromotive force and casting it in Maxwell-Lorentz-Einstein electromagnetism.

\section{Electromagnetic induction within Maxwell - Lorentz - Einstein Electromagnetism}\label{mle}
In an axiomatic presentation of Maxwell-Lorentz-Einstein (MLE) Electromagnetism, one usually
begins with Maxwell's equations in vacuum (in the vectorial form given
to them by Oliver Heaviside \cite{oliverh}.)
If we define the charge density $\rho$ and the current density $\vec J$, the electric and the magnetic fields $\vec E$ and $\vec B$, together with the two constants $\varepsilon_0$ and $\mu_0$ remain still without physical dimensions   and physical meaning.
The assumption of the Lorentz force:
 \begin{equation}\label{lorforce}
    \vec F= q(\vec E+ \vec v\times \vec B),
     \end{equation}
      allows to establish the physical dimensions of the two fields $\vec E$ and $\vec B$, together with those of the two
 constants $\varepsilon_0$ and $\mu_0$.
     \par
     Following \cite{epl}, we shall define  the {\em induced}
electromotive force as:
\begin{equation}\label{natural}
   \mathcal E = \oint_l (\vec E + \vec v_c \times \vec B)\cdot \vec{dl},
\end{equation}
where $\vec v_c$ is the velocity of the positive charges which are, by convention,
the current's carriers, and $\vec E, \vec B$ are
solutions of Maxwell's equations.
This integral yields, numerically,  the work done by the Lorentz force
 on a unit positive point charge along the considered
closed path. Of course, the definition (\ref{natural}) is an assumption whose validity rests on the experimental corroboration of all predictions derived from it.
Since
\begin{equation}\label{vecpot}
\vec E = - \nabla\varphi -{{\partial \vec A}\over{\partial t}},
\end{equation}
with $\varphi$ the scalar potential and $\vec A$ the vector potential, we have:
\begin{equation}\label{natural2}
   \mathcal E = \oint_l \left[\left(-\nabla\varphi- \frac{\partial \vec A}{\partial t}\right) + (\vec v_c \times \vec B)\right]\cdot \vec{dl}=
\oint_l \left[\left(- \frac{\partial \vec A}{\partial t}\right) + (\vec v_c \times \vec B)\right]\cdot \vec{dl}.
\end{equation}
Equations (\ref{natural}, \ref{natural2}) are  the same as Maxwell's
  with  the fundamental specification that the velocity $\vec v_c$ is the velocity
of the positive charges and not the velocity of the circuit element containing them.
  They are
valid for any integration line.
\par
We attribute an electromotive force \cite{names} also to batteries or sources
of alternate current.  A general definition of what an electromotive force is, was given, for instance,  by Slater and Frank in their concise textbook, {\em Electromagnetism}:
\begin{quote}\small
    By definition, the {\em emf} around a circuit equals the total work done, both by electric and
magnetic forces and by any other sort of forces, such as those concerned in chemical
processes, per unit charge, in carrying a charge around the circuit \cite[p. 79]{sf}.
\end{quote}
      Equations (\ref{natural}, \ref{natural2}) are  local laws \cite{causality}:   they relate the line integral quantity $\mathcal E$ at the instant $t$
      to other physical
   quantities defined at each point of the integration line at the same instant $t$. In the case of a rigid
   filiform circuit, equations (\ref{natural}, \ref{natural2}) are Lorentz-invariant \cite{epl} and Appendix \ref{inertial}).
  A proper
   description of the relative inertial motion of a magnet and a rigid  conducting loop has been considered by Einstein as one
    of the reasons for the development of special relativity \cite{ein05r}. The proof of Lorentz
invariance already given in \cite{epl} concerns the relative motion
 of a magnet and a circuit. It is easy to extend the proof to the general case in which the source of the magnetic
 field is not specified (Appendix
 \ref{inertial}).
 \par
 Equation  (\ref{natural2}) implies that there are two independent contributions to the induced {\em emf}: the time variation
 of the vector potential  and
  the effect of the magnetic field on  moving charges.
 If  every element of the circuit is {\em at rest}, $\vec v_c=\vec v_d$, where $\vec v_d$ is the drift velocity of the (positive) charges.
 Then, equation (\ref{natural2}) assumes the form:
 \begin{equation}\label{leggedrift}
    \mathcal E= -\oint_l \frac{\partial \vec A}{\partial t}\cdot \vec
dl
    +\oint_l (\vec v_{d}\times \vec B)\cdot\vec {dl}.
\end{equation}
This equation shows that, in general, the drift velocity contributes
to the induced {\em emf}. In filiform circuits, the second line
integral is null because, in every line element,  $\vec v_d$ is
parallel to $\vec {dl}$. In extended conductors, the drift
velocity plays a fundamental role. As shown in \cite{epl},  the case of Corbino's disc is particularly interesting since
the application of Eq. (\ref{leggedrift}) explains the
magnetoresistance effect without the need of microscopic
 models.
  \section{Where is the induced electromotive force  localized?}\label{where}
\begin{figure}[h]
\centering{
\includegraphics[width=7cm]{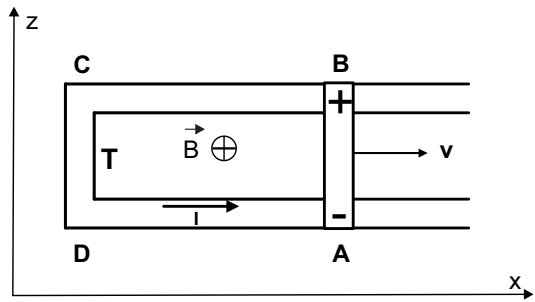}\\
}
 \caption{\label{barra}
A bar AB of length $a$ moves with constant velocity $\vec v$ along the U-shaped frame T
  in a uniform and constant magnetic field $\vec B$, produced by a source at rest in the laboratory and directed along the positive direction of the $y$ axis. The arrow in the low part of the frame indicates the direction of flow of the current $I$. }
\end{figure}

As a consequence of the widespread adoption of the ``flux rule'', the issue of the spatial localization of the induced {\em emf} is usually ignored. Indeed, as we will show below, the \fr cannot say where the induced {\em emf} is localized. Hence, for the users of the \fr, the question is meaningless.
 However, we shall show that this question indeed
has physical meaning by considering the illustrative
example of a bar moving along a U-shaped conducting frame and immersed in a uniform and constant magnetic field (fig. \ref{barra}).
  \par\noindent
This thought  experiment   goes back to
 Maxwell \cite[pp. 218-219]{treatise2}. Conceptually,  it can be considered as a
version of Faraday's disc \cite{epl}  in which the rotation of the disc is
substituted by the inertial motion of part of the circuit, thus
allowing the use of two inertial reference frames. In the following discussion, it is conventionally assumed that the mobile charges are positive.   Because in the
laboratory reference frame the vector potential is constant, the
induced {\em emf} is due only to the line integral $\oint_l (\vec v_c\times\vec B)\cdot \vec{dl}$ of Eq. (\ref{natural2}), and the only portion of the circuit in which it is non-zero is the AB segment. Therefore, the induced {\em emf} is given by $\mathcal E= -B_yva$ (the circulation along the contour is counted as positive when clockwise in fig. \ref{barra}). The
{\em emf} is localized in the bar: the bar acts as a battery, with the
positive pole at point $B$. The current flows from  $A$ to $B$, i.e., from the
point at lower potential to that at higher potential, like in a battery.
However, for a pair of points on the immobile frame T, the current flows
from the point at higher potential to that at lower potential. The
concept of localization of the induced {\em emf} therefore has physical
meaning because it allows the prediction -- testable by experiment
--  of how the potential difference between two points is related to
the  current flow.  If we drop this concept,
we decrease the predictive and explanatory power of the theory.
\par
 The \fr
  predicts the correct induced {\em emf}. The flux of the magnetic field through the area
  ABCD is  $\Phi= B_yavt$, if at  $t=0$ the bar AB coincides with the arm DC of the frame.
  Then:
  \begin{equation}\label{emfbar}
     \mathcal E= - \frac{d\Phi}{dt}=-B_yav,
  \end{equation}
   where the minus sign indicates that the induced current circulates counterclockwise, as
  indicated in fig. \ref{barra}.   The ``flux rule''  cannot say where the {\em emf} is localized; it can only guess that it might  be   localized in the bar AB because it is moving (on the grounds  that it is the bar's motion that produces the variation of the    magnetic field's flux  and, hence, the induced {\em emf}).
   \par
   However, applying the same argument in the (primed) reference frame of the bar AB, it will be said that
    the {\em emf} is localized in the opposite vertical arm CD of the frame because it is seeing moving away from the bar along the negative direction of the common $x\equiv x'$ axis. This statement is false, because,
    in the reference frame of the bar AB also, the {\em emf} is localized in the bar.
In fact, at every point of the bar's reference frame, there is an electric field --
due to the fields' transformation equations (see Appendix \ref{inertial}) --  given by:
\begin{equation}\label{elezeta}
  E_z'=\Gamma  vB_y.
\end{equation}
Therefore,  an {\em emf} $\mathcal E'=\Gamma vB_y
a=\Gamma \mathcal E$ is induced in the bar.
 In the arm CD, the effect of the electric field $E_z'$ on the charges is exactly balanced by the
 magnetic component of the Lorentz force, and due  to the motion ($-v$) of the arm CD
 in the field $B'_y=\Gamma B_y$.
The relation $\mathcal E'=\Gamma\mathcal E$, thus obtained,  is
a particular case of the more general one treated in  Appendix
\ref{inertial}.
 \begin{table}[h]
  \centering{
  \begin{tabular}{|l|c|c|c|}
\hline
\bf {Electric field}& \bf{Symbol} & \bf{Value} &\bf{Direction} \\
\hline  \hline
Induced & $E_i$ & $vB$ &$\uparrow$ \\
\hline
Effective &$E_e$&$vBa/l$ & $\uparrow$ \\
\hline
 Electrostatic  &$E_{es}$& $vB(1-a/l)$& $\downarrow$ \\
\hline
\end{tabular}
  \caption{Electric fields in the bar. The arrow indicate the direction of the field along the $z$ axis.}\label{eleinbar}
  }
\end{table}
\par\noindent
 If the entire circuit is
homogeneous (the frame and the bar are of the same material and
have the same section), the physics of the circuit is as follows (see Table \ref{eleinbar}). At
every point of the circuit (bar included), there is a current density given by:
\begin{equation}\label{effective}
J=\frac{E_e}{\rho}=\frac{1}{\rho}\frac{vBa}{l},
\end{equation}
 where $E_e$ is an ``effective electric field'', $\rho$ the
resistivity of the material, and $l$ the length of the entire circuit
(frame + bar). In the bar, the effective electric field $vBa/l$ is the sum of
two opposing fields: the induced electric field $\vec
E_i=vB\hat z$ and an electrostatic field $E_{es}=E_i- E_e$ directed
along the negative direction of the $z$ axis.
The origin of this field
can be explained as follows. Let us consider the bar moving
without touching the frame\label{3fields}. The induced electric
field $E_i =vB$, pushes the positive charges towards the
point $B$: this point becomes positively charged, and point $A$
negatively charged. Therefore, an electrostatic field $E_{es}$
directed from $B$ to $A$ is established:
this field nullifies the induced electric field in a stationary condition.If the bar slides while touching the frame, a current flows in the circuit: while the
induced electric field maintains its original value, the value of the electrostatic
 field $E_{es}$ diminishes and is such that $E_{es}=E_i-E_e$ in stationary regime.
 Finally, it can be easily verified that the field defined
as $\vec E_{c}= \vec E_e$ in the frame and as $\vec
E_{c}=\vec E_{es}$ in the moving bar is conservative. This result allows calculating the potential difference between two arbitrary points of the circuit, which is given by the line integral of the conservative electric field between the two points. Naturally, the calculation leads to Ohm's law, as implied by the starting equation (\ref{effective}), which is Ohm's law in local form.
\par
 The bar we have discussed acts precisely like a battery. The circuital laws are, therefore, similar. We know that complicated chemical reactions take place in a battery. However, its circuital behavior can be described with a model which involves similar $E_i$, $E_e$, and $E_{es}$. The induced electric field of the bar corresponds to the electromotive field of the battery; the electrostatic field is present in both cases, and in both cases, the effective electric field inside the battery is given by the difference between the previous two.
\par
One must supply mechanical power to keep the bar AB in motion with a constant velocity. If the bars motion is frictionless,this power is equal to the electrical power dissipated
in the circuit.
Indeed,  the electrical power supplied by the electromotive
force  is given by\label{see}:
\begin{equation}\label{work}
    W(\mathcal E)= \mathcal E I=\frac{\mathcal E^2}{R}=\frac{(vBa)^2}{R},
\end{equation}
where $R$ is the resistance of the circuit. This power is dissipated as heat. For energy conservation, the
power supplied by the electromotive force must come from the
mechanical work done by the force  keeping the
bar in motion. If a current $I$ flows in the bar along the positive direction of the $z$ axis, the magnetic field $B$ exerts a force on the bar given by $-IBa\hat{x}$: it tends to slow down the bar. An equal and opposite force must be supplied to the bar to keep it in motion. Its magnitude is given by:
\begin{equation}\label{forceon wire}
F= IBa=\left(\frac{vBa}{R}\right)Ba.
\end{equation}
The mechanical power necessary for maintaining the bar in
motion is:
\begin{equation}\label{mechpower}
W(F)= Fv=\frac{(vBa)^2}{R}=\frac{\mathcal E^2}{R}=W(\mathcal E).
\end{equation}
 The fact that the magnetic field $B$ appears in the expression of
the electrical power supplied by the induced {\em emf}
contrasts the fact that, in a vacuum, the magnetic force does not
produce work on a moving charge.
Here, the magnetic field plays only the role of a mediator between the mechanical force and the electrical power (through the term $\vec v_c\times \vec B$) without yielding any
amount of energy.
\par
Let us note that the system described here is quasi-stationary because the length of the frame entering the circuit increases with time.
\section{Why the  ``flux rule'' is not a physical law}\label{rulesec}
 In this section, we shall compare  the general law of electromagnetic induction  (\ref{natural2}), derived within MLE Electromagnetism, with the \fr (\ref{flusso}).
  Note that the expression of the induced {\em emf} (\ref{natural2}) contains the vector potential $\vec A$, as in Maxwell's formula (\ref{leggegenmax}).
\par
We shall consider a
 closed conducting filiform loop immersed in a magnetic field and in motion. This loop is not assumed to be rigid, so that it can deform. As is usual, the induced {\em emf} can also be written in terms of the magnetic field.
Starting from
Eq. (\ref{natural}), we write, in the reference frame of the laboratory:
\begin{eqnarray}\label{onlyone}
  \mathcal E=  \oint _{l}^{}{\vec E\,\cdot\, \vec{dl}} + \oint_l (\vec v_c\times\vec B) \cdot \vec{dl}&= &\int_{S} \nabla \times \vec E \,
\cdot \, \hat n \, dS + \oint_l (\vec v_c\times\vec B) \cdot \vec{dl}\nonumber\\
&&\\
&= &- \int_{S} \frac{\partial \vec B}{\partial t} \, \cdot \, \hat n \, dS + \oint_l (\vec v_c\times\vec B) \cdot \vec{dl}\nonumber,
\end{eqnarray}
where $S$ is any arbitrary surface that has the closed wire $l$ as contour.
We then use the identity, valid for every vector field with null divergence (see, for instance, \cite[pp. 10 - 11]{andy}:
\begin{equation}\label{identity}
  \int_S \frac{\partial \vec B}{\partial t}\cdot  \hat n \,dS=\frac{d}{dt}\int_S \vec B \cdot \hat n \,dS+\oint_l (\vec v_l\times \vec B)\cdot\vec{dl},
\end{equation}
where $\vec v_l$,  the velocity of the wire element $dl$,  can be different for each wire element.
Then,  Eq. (\ref{onlyone}) becomes:
\begin{equation}\label{quasiflusso}
      {\mathcal E}=- \frac{d\Phi}{dt}- \oint_l(\vec v_l\times \vec B)\cdot\vec{dl} +\oint_l (\vec v_c \times \vec B)\cdot \vec{dl}.
\end{equation}
In the case of a {\em rigid, filiform  wire} moving with velocity $V$
along the positive direction of the common $x'\equiv x$ axis, this
equation takes the form:
\begin{equation}\label{quasiflussorigido}
{\mathcal E}=- \frac{d\Phi}{dt}- \oint_l(\vec V\times \vec B)\cdot\vec{dl} +\oint_l (\vec v_c \times \vec B)\cdot \vec{dl}.
\end{equation}
This equation is not the \fr. To get it, we must abandon special relativity and try using Galilean relativity. Since $ V\ll c$ and $ v_d\ll c$, we can  write:
$\vec v_c= \vec V +\vec v_d$ ($c=\infty)$.
Then, we get:
\begin{equation}\label{quasiflux2}
\mathcal E=-\frac{d\Phi}{dt}
+\oint_l (\vec v_{d}\times \vec B)\cdot\vec dl= -\frac{d\Phi}{dt},
\end{equation}
i.e. the ``flux rule'' (the line integral is null because, for every wire
element, the drift velocity of the charges $\vec v_d$ is parallel to $\vec {dl}$).
\par
In the reference frame of the circuit, we have:
\begin{equation}\label{emirelcirc}
    \mathcal E'= - \int_{S'} \frac{\partial \vec B'}{\partial t'} \, \cdot \, \hat n' \, dS' + \oint_{l'} (\vec v'_d\times\vec B') \cdot \vec{dl'}=- \frac{d}{dt'}\int_{S'} \vec B' \cdot\hat n' dS'=-\frac{d\Phi'}{dt'}.
\end{equation}
The ``flux rule''  -- for filiform and rigid circuits --  is valid in both reference frames if one
uses the Galilean transformation of velocities, namely Galilean relativity ($c=\infty$). Notice that using the Galilean velocity composition is a sufficient condition for the \fr being Galileo-invariant. It is also necessary because, without the Galilean approximation, the \fr cannot even be established: we would have to stop at equation (\ref{quasiflussorigido}).
  Finally, notice that the \fr is Galileo-invariant at first sight. In the Galilean limit ($c=\infty$) of the relativistic transformations of electric and magnetic fields, $\vec B'=\vec B$. On the other hand, areas and time intervals have the same value in two inertial reference frames. It follows that $d\Phi/dt=d\Phi'/dt'$, i.e., that the \fr is Galileo invariant and the induced {\em emf} has the same value in every inertial frame. How can a Galileo-invariant equation be accepted in a relativistic theory as MLE electromagnetism? These considerations constitute a first argument against the "flux rule" being a physical law.
\par
  Second, let us note that the integral in Eq. (\ref{quasiflux2}) is null only if the circuit is filiform. If not, the extension of the material makes $\vec {v}_d$ non parallel to the circuit element $\vec {dl}$, and the "flux rule" is not obeyed.
    In fact:
\begin{equation}\label{quasiflux}
\mathcal E=-\frac{d\Phi}{dt}  +\oint_l (\vec v_{d}\times \vec B)\cdot\vec {dl}.
\end{equation}
Differently from equation (\ref{quasiflux2}), the integral is not null owing to the extension of the material.
\par
Third, we would like to emphasize that the general law (\ref{natural2}) is a local one.
However, the ``flux rule'' does not satisfy the
locality condition. It relates the {\em emf} induced in a
conducting filiform circuit at time $t$ to the time variation -- at the exact same instant $t$ -- of the flux of the magnetic
field through an {\em arbitrary} surface that has the circuit as a contour.
Consequently, the ``flux rule'' cannot be causally interpreted, because what happens at the surface at time $t$ cannot influence what
happens in the circuit at the same instant $t$, unless physical interactions can propagate with infinite
speed. Moreover, since one can arbitrarily choose the integration surface,  we should have endless causes of the same effect:  the variation of the flux of the magnetic field through any arbitrary surface, however large, which has
   the wire as its contour would then be the {\em cause} of the induced {\em emf}, thus violating, again, the locality condition.  As shown above in the case of the moving bar,  the ``flux
rule'', though predicting the correct value of the induced {\em emf}, cannot say anything about
   the physical processes involved or about their causal nature. It is just a mathematical relation between two physical quantities, devoid of
   physical insight. It can be used, with great care, as a calculation shortcut only when guided by the predictions of the general law.
\par
Moreover, the ``flux rule'' does not always yield the correct prediction.   Indeed, in 1914, Andr\'e Blondel showed that there could be a flux variation without any induced {\em emf}, thus falsifying the \fr  \cite{blondel}.
\begin{figure}[h]
\centering{
\includegraphics[width=3cm]{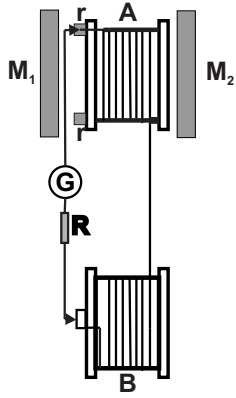}\\
}
  \caption {Sketch of Blondel's experiment testing the \fr. $M_1$ and $M_2$ are the circular plates of an electromagnet: the magnetic field between the two plates is equal to $0.08$ T and is `reasonable uniform'. R is a resistance, and G is a D'Ansorval galvanometer. A and  B are two wooden cylinders. The two ``r''  represent the intersection of a conducting ring attached to cylinder A with the plane of the figure. The two arrows indicate sliding electric contacts. A conducting coil wound around cylinder A, placed within the electromagnet, is unrolled and transferred to cylinder B, where the magnetic field is null.  }\label{blondel}
\end{figure}
\par\noindent
  Blondel used a solenoid wound around a wooden cylinder  A placed between the circular armature plates of an electromagnet (Fig. \ref{blondel}). He attached one end of the solenoid to another parallel wooden cylinder B outside the electromagnet (where the magnetic field is null). He then unrolled the solenoid as he  transferred the coils from cylinder A to  cylinder B, while maintaining the unrolling wire tangent to the two cylinders \cite{blondel}, \cite{ggvp}. According to the \fr, during the transfer  of the coil from cylinder A to cylinder B, an {\em emf} should be induced in the unrolling solenoid, according to the equation:
\begin{equation}\label{emfblondel}
|\mathcal E| =\left|-\Phi_{\rm single\,coil}\frac{dN}{dt}\right|,
\end{equation}
where $\Phi_{\rm single\,coil}$ is the magnetic flux through one coil and $N$ the number of coils on cylinder $A$.
Since the radius of a coil was $0.1$ m, $B = 0.08$ T and $|dN/dt| = 6.66$, the {\em emf} predicted
by (\ref{emfblondel}) is $0.017$ V, well above the detection capability of the D'Ansorval's galvanometer Blondel used to monitor the current in the circuit. Contrary to what is predicted by the "flux rule", the galvanometer showed no deviation \cite{blondel, ggvp}. Of course, the general law (\ref{natural2}) predicts that in Blondel's experiment, there is no induced {\em emf}. Indeed, the first line integral is null because the vector potential does not depend on time. If, during the unrolling of the solenoid, the unrolling wire is kept tangent to the two cylinders,  the second integral is also null because the term $\vec v_c\times \vec B$ has a null component along the wire.
\par
Blondel's experiment has been rapidly forgotten \cite{rotore}.
\section{Why did Maxwell's general law fall into oblivion?}\label{why}
We have already stressed that the abstract and complicated derivation of Maxwell's general law may have been an obstacle to its acceptance for his contemporaries (section \ref{maxind}).
  The elimination of the vector potential by Hertz and Heaviside from Maxwell's Electromagnetism has very likely contributed to the oblivion of Maxwell's general law \cite{ggvp}.   In particular, one aspect might have had a lasting conceptual impact. Heaviside doubted that the vector potential could have physical meaning: ``\dots  [Maxwell] makes use of an auxiliary function,
the vector potential of the electric current, and this rather complicates the matter, especially
as regards the physical meaning of the process. It is always desirable when possible to keep
as near as one can to first principles'' \cite[p. 46]{heavi}. Hertz compared the vector potential to a scaffolding that can be removed after the building has been completed. Hertz too was doubtful about the physical meaning of the vector potential: ``\dots one would expect to find in these [fundamental]
equations relations between the physical magnitudes which are actually observed, and not
between magnitudes which serve for calculations only'' \cite{ew}.
 \par
 However, other factors may have contributed, and in particular the role played by textbooks.
 In his  {\em The Structure of Scientific Revolutions}, Thomas Kuhn has stressed the primary role played by textbooks
in transmitting the acquired knowledge to new generations \cite{kuhn}.
To verify the role played by textbooks in the oblivion of Maxwell's general law, we have analyzed eleven university textbooks -- considered representative due to their authors or their popularity.
Here is a concise overview; one can find the detailed analysis in \cite{SM}.
\par
 In all these textbooks, the treatment of induction begins with the presentation of a series of Faraday's experiments, distinguishing between those involving the relative motion of a magnet and a circuit (magneto-electric induction, in Faraday's language) and those in which the induced current is due to the current variation in another circuit (volta-electric induction).
 The theoretical description of these phenomena varies greatly, ranging from the elementary treatment by Riecke \cite{riecke}, to the more sophisticated ones involving the relativistic nature of Electromagnetism \cite{landau_ecm, jackson}.
\par
Except for Feynman's {\em Lectures} \cite{feyn2}, Zangwill's \cite{andy}, and Griffiths' \cite{griff}] texts, all these textbooks  refer to the ``flux rule'' as the
``law of electromagnetic induction''.  These texts adhere, knowingly or not, to the introductory part of Maxwell's
discussion of electromagnetic induction and ignore his general law.
\par
    Indeed, a typical theoretical treatment  (see for instance \cite[p. 233]{jackson}) is based on the definition of the induced {\em emf} as:
\begin{equation}\label{emiC}
\mathcal E_C=  \oint_{l}\vec E\cdot \vec {dl}.
\end{equation}
  This definition stems from the electrostatic law $\vec F = q\vec E$, and
          it leads to the correct result only if applied in the reference frame of a rigid filiform circuit. It cannot be used when the circuit moves.
\par
By using  Stokes' theorem and  Maxwell's equation:
\begin{equation}\label{thirdmax}
\nabla\times \vec E=-\frac{\partial\vec B}{\partial t},
\end{equation}
it is found that:
\begin{equation}\label{howtofr}
\mathcal E_C=  \oint_{l}\vec E\cdot \vec {dl}=\int_S \nabla\times \vec E\cdot \hat n\, dS=
-\int_S \frac{\partial\vec B}{\partial t}\cdot \hat n\, dS=-\frac{d}{dt}\int_S\vec B\cdot\hat n\, dS,
\end{equation}
namely, the \fr. Notice that the last equality is valid only if the filiform and rigid circuit is at rest.
\par
Another type of approach consists in stating the ``flux rule'' as a law inferred from experiments -- as Maxwell did as a first step --  and exploring how it effectively deals with many experimental situations.
Within this approach, starting from the ``flux rule'', the differential Maxwell's equation of the {\em curl}
of the electric field is derived \cite[p. 80]{sf}; \cite[pp. 158-159]{pp}; \cite[pp. 158-159]{purmor}.
\par
 Modern textbooks widely use the vector potential to calculate the effects of currents or magnets. Nevertheless, they do not use the vector potential to treat electromagnetic induction: see, for instance, \cite[pp. 462-463]{andy}.
This choice implicitly does not make any  distinction between equations that obey the locality condition and those that don't.
The general law of electromagnetic induction belongs to the former; the ``flux rule'' to the latter.

\section{Teaching issues}\label{teaching}
Teaching the modern formulation of Maxwell's
general law in advanced courses poses no problems. However, it may be more difficult in high schools or elementary physics courses \cite{high_school}.  Here, we suggest a way of introducing Maxwell's general law and
  the vector potential by using Faraday's well-known experiment, commonly reproduced in
  didactic laboratories:   the relative motion of a magnet and a rigid, filiform coil (see Appendix \ref{inertial}). \cite{labexp} describes a laboratory session in which, starting from the autonomous experimenting by students, the general law of electromagnetic induction is obtained through a collective discussion that is guided, when necessary, by the instructor.
  \par
  The starting point for the theoretical description of the (thought) experiment should be the definition of the induced {\em emf} given by Eq. (\ref{natural}).    Instructors should adapt the formalism to their
  teaching contexts, while keeping  two cornerstones:  the necessity of describing the phenomenon
   in the two reference frames and that of a {\em local} description in the coil's reference frame.
      \par
     Eq. (\ref{natural2}) is written step by step by asserting that a physical law must have the same form in every inertial frame. Our proposal derives the expression of the electric field entering the definition (\ref{natural}),  thus obtaining the modern formulation of Maxwell's general law. It must be stressed that Eq. (\ref{natural2}) will not be derived but only reasonably guessed.
       \par
   In the reference frame of the magnet, the magnetic field is constant, and there is no electric field.Therefore, the induced {\em emf} can be due only  to a term $\vec V \times \vec B$,
   as suggested by  the magnetic component
   of the Lorentz force ($\vec V$ is the velocity of the coil with respect to the magnet):
\begin{equation}\label{magnethigh}
    \mathcal E=\oint_l(\vec V\times\vec B)\cdot\vec{dl}.
\end{equation}
       In the reference
   frame of the coil,  the induced current must be due to an electric field $\vec E'$ such that:
\begin{equation}\label{ecoil}
    \mathcal E'=\oint_{l'}\vec E'\cdot \vec{dl'}=- \frac{d}{dt'}\int_{S'}\vec B'\cdot \hat n' dS'.
    \end{equation}
Following Maxwell \cite[pp. 27-28]{treatise2} ),  we
write this equation in terms of a vector defined at each point of the
coil: this will ensure the locality of the phenomenon.
Let us denote this vector  $\vec A'$ and call it ``vector
potential'':
\begin{equation}\label{potvect_hs}
   \mathcal E'=\oint_{l'}\vec E'\cdot \vec{dl'}=- \frac{d}{dt'}\int_{S'}\vec B'\cdot \hat n' dS'=-\frac{d}{dt'}\oint_{l'}\vec A'\cdot\vec{dl'}.
\end{equation}
Naturally, this is Maxwell's Eq. (\ref{emfdef}). Because the integration line
does not depend on time, the above equation assumes the form:
\begin{equation}\label{potvect_hs2}
   \mathcal E'=\oint_{l'}\vec E'\cdot \vec{dl'} =\oint_{l'}\left(-\frac{\partial \vec A'}{\partial t'}\cdot\vec{dl'}\right).
\end{equation}
Because a physical law must be valid in every
inertial frame, we conclude that the law of electromagnetic
induction should be written as a combination of Eq.
(\ref{magnethigh}) and (\ref{potvect_hs2}), i.e. as:
\begin{equation}\label{conclusion}
    \mathcal E'= \oint_{l'} \left[ -\frac{\partial \vec A'}{\partial t'}+ (\vec V\times \vec B')\right]\cdot \vec {dl'}.
\end{equation}
This equation is valid in both reference frames: the second
term of the integral operates in the magnet's reference frame; the
other in the coil's frame.
At this point, one should recall that Eq. (\ref{conclusion}) is valid also if, instead of a moving magnet, we have a primary circuit in which a variable current flows: see Appendix \ref{inertial} and the discussion of  Purcell and Morin's textbook in \cite{SM}.
\par
 From this point on, instructors can refine the proposal according to their teaching contexts, having in
mind the complete treatment of section \ref{mle}. One
could introduce the distinction between the velocity of the coil and
that of the charges contained in it. As for the expression between
the two square brackets of Eq. (\ref{conclusion}), the instructor could
generalize it by adding the term $-\nabla \varphi$.
 \begin{figure}[h]
  \centering
  \includegraphics[width=6cm]{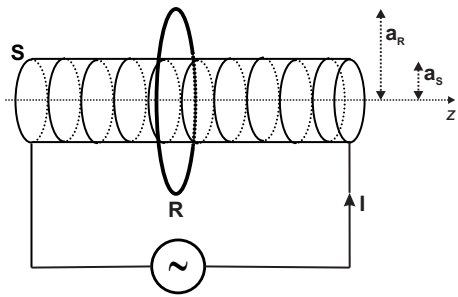}\\
  \caption{Induction between a ring $R$ and a solenoid $S$ through which flows a low frequency alternating current $I$. See also \cite{ggvp, rousseaux}}.\label{ml}
\end{figure}
\par\noindent
By considering the apparatus of fig. \ref{ml}, one can derive the
expression of the vector potential in an exemplar case. A low frequency alternating current $I$ flows in a long solenoid S of radius $a_S$, thus creating a time-varying magnetic field. In the case of an ideal solenoid (constituted by an infinite array of circular coils), the magnetic field outside the solenoid is null, whereas $\vec B=\mu_0nI\hat{z}$ inside the solenoid, with $n$ the number of turns per unit length of the solenoid, and $\hat{z}$ the unit vector along the solenoid's symmetry axis.
\par
According to the  law  (\ref{conclusion}), the {\em emf} induced in
the filiform ring is given by:
\begin{equation}\label{anello}
    \mathcal E = -\oint_R \frac{\partial \vec A}{\partial t} \cdot
\vec{dl}=
    -\frac{d}{dt}\oint_R \vec A\cdot \vec {dl}= - 2\pi a_R \frac{dA}{dt},
\end{equation}
where $a_R$ is the radius of the ring.  The induced {\em emf} can also
be written as:
 \begin{equation}\label{anello_regola}
    \mathcal E =-\frac{d}{dt}\int_{\mathcal S} \vec B\cdot\hat n\, d\mathcal S=-\mu_0n\pi a_S^2\frac{dI}{dt},
 \end{equation}
 where ${\mathcal S}$ is an arbitrary surface that has the ring as contour.
  By equating the last members of the  two equations above, we get:
 \begin{equation}\label{potvett}
    A=\mu_0\frac{n a_S^2 }{2a_R}I,
 \end{equation}
 where the constant of integration is taken to be to zero to obtain a vector field that vanishes far from the source.
   The vector potential has the same absolute value at every point of
  the ring and the induced {\em emf} is distributed homogeneously along the ring: the ring acts as a source of current.  Taking two
  points $M$ and $N$ on the ring, we have the electric potential difference:
  \begin{equation}\label{ringequi}
    \varphi_M-\varphi_N = \mathcal E\frac{\Delta \alpha}{2\pi}- i \Delta R=\mathcal E\frac{\Delta \alpha}{2\pi}-\frac{\mathcal E}{R}
    \left(R\frac{\Delta \alpha}{2\pi}\right)=0,
  \end{equation}
where $i$ is the induced current, $\Delta R$ the resistance of the $MN$ arc  and
$\Delta\alpha$ its  angle.
The filiform ring is an equipotential line.
\par
Eq. (\ref{potvett}) shows that the sources of the vector
potential are the currents -- namely charges in motion -- as it is  for
the magnetic field. Eq. (\ref{potvect_hs2})
shows that the vector potential (i.e., the charges in motion) contributes
to the value of the electric field through its partial derivative with
respect to time. On the other hand, Eq. (\ref{potvect_hs})
establishes a relation between the vector potential and the
magnetic field, {\em at the same instant}, in different regions of
space: a line and an arbitrary surface that has the line as a contour.
The arbitrariness of the choice of the integration
surface suggests a strong spatial relationship between the vector
potential and the magnetic field:  instructors know that $\vec B=
\nabla\times\vec A$ and, depending on the teaching context,
can communicate to the students the explicit form of this equation
or its essential feature (dependence of the magnetic field on the
spatial partial derivatives of the vector potential components).
In other words, the vector potential depends both on the spatial
coordinates and on time. The dependence on the coordinates
relates the vector potential to the magnetic field; the dependence
on time, relates the vector potential to the electric field.
\par
 To wind it all up, the instructor should discuss again the issue of
the transition from the conceptual framework of the action at a distance to a
field description of electromagnetic phenomena. In addition to the path going from the existence of a charge to the explanation of the force between two charges (charge $\rightarrow$ fields
$\rightarrow$ force on another charge), there is another one
according to which charges create potentials that give rise to fields and, in turn, to a force exerted on other charges (charge $\rightarrow$ potentials $\rightarrow$
fields $\rightarrow$ force on another charge). Here, we have an example
of how physicists use mathematics to build increasingly abstract theories, but which are, sometimes -- as in the case of the
potentials -- more apt at dealing with the intrinsic Lorentz
invariance of Maxwell's electromagnetism.
\section{Conclusions}
 The rooting and the persistence of the \fr appear to be an example of the collective processes studied by Ludwik Fleck in his {\em Genesis and Development of a Scientific Fact}, first published in German in 1935 \cite[p. 27]{fleck}]:
  \begin{quote}\small
 Once a structurally complete and closed system of opinions consisting
of many details and relations has been formed, it offers
enduring resistance to anything that contradicts it.
\par
[\dots]
\par
What we are faced with here is not so much
simple passivity or mistrust of new ideas as an active approach
which can be divided into several stages. (1) A contradiction to the
system appears unthinkable. (2) What does not fit into the system
remains unseen; (3) alternatively, if it is noticed, either it is kept
secret, or (4) laborious efforts are made to explain an exception in
terms that do not contradict the system. (5) Despite the legitimate
claims of contradictory views, one tends to see, describe, or even
illustrate those circumstances which corroborate current views and
thereby give them substance.
 \end{quote}
 In our opinion, this description fits to a large extent the case of the \fr.
Textbooks played a fundamental role in this process. Understanding this role and bringing to the surface its main traits would require a study encompassing physics, epistemology, and social behaviors. It goes
beyond the aims of this paper and our competence. We can only underline some aspects of the issue. Generally, a rooted theoretical description of some phenomena is shaken when a new experimental result contrasts the accepted view. In the case of the \fr, Blondel's experiment should have put it under fire. However, this did not happen. Thomas Kuhn has described the resistance of an acquired view against experimental results that undermine it. Generally, the change of an acquired view is provoked by a series of new experimental results and/or by a radical change in a broader theoretical contest. Einstein's relativistic electromagnetism should have played this role. Nevertheless, it did not. Indeed, textbooks sometimes touch on this issue but do not wholly develop its consequences to the point of asking: what invariancy is satisfied by the \fr? Moreover, if it emerges that the \fr is Galileo-invariant, how can it be tolerated in a wholly relativistic theory?
 Another critical issue is constituted by the historical falsehoods encountered in many texts. These falsehoods are possible because textbook writers do not check their statements with historical records but adhere to some anonymous historical tradition. This habit amounts to dismissing the cultural role of historical studies. Historical falsehoods can also distort physics. This issue should also be explored from a sociological point of view.
 Finally, it seems clear that the lack of an epistemological commitment favors the tradition's persistence.

 \vskip3mm\par\noindent
{\bf Acknowledgements.} I want to thank Biagio Buonaura for the many helpful discussions on this subject.
The comments and suggestions of the two anonymous Referees have also contributed to giving the paper its final form.

\appendix
\small
\section{Lorentz invariance of the general law of electromagnetic induction}\label{inertial}
Let us consider a magnetic field source at rest in the laboratory reference frame.  A filiform, rigid circuit moves with velocity $V$ along the positive direction of the common $x\equiv x'$ axis.
In the circuit's reference frame, the general law (\ref{natural2}) assumes the form:
 \begin{equation}
{\mathcal E'} =    \oint{\vec E'\cdot \vec{dl'}}+ \oint(\vec v'_d \times \vec B')\cdot\vec{dl'},
\end{equation}
where $\vec v_d'$ is the drift velocity of the charges.
The second integral is null because the drift velocity is parallel to $\vec {dl'}$ in every circuit element, so that we are left with the first line integral.
Taking into account the field's transformation:
 \begin{eqnarray}
   E'_x&=& E_x\nonumber\\
   E'_y&=& \Gamma[E_y+ (\vec V\times \vec B)_y]\nonumber\\
    E'_z&=& \Gamma [E_z +(\vec V\times \vec B)_z]\nonumber,
\end{eqnarray}
where $\Gamma=1/\sqrt{1-V^2/c^2}$,
and  the coordinates transformations:
\begin{eqnarray}
dx'&=&\Gamma dx\nonumber\\
dy'&=&dy\nonumber\\
 dz'&=&dz\nonumber,
\end{eqnarray}
we get:
\begin{eqnarray}\label{invapp}
    \mathcal E'&=& \Gamma\oint E_xdx+\Gamma \oint [E_y+(\vec V\times\vec B)_y]dy+\Gamma \oint [E_z+(\vec V\times\vec B)_z]dz\nonumber\\
&&\\
&=&\Gamma \oint[\vec E+(\vec V\times\vec B)\cdot\vec{dl}]\nonumber.
\end{eqnarray}
In the laboratory reference frame, we have:
\begin{equation}\label{appemilab}
    \mathcal E= \oint (\vec E + \vec V\times\vec B)\cdot \vec {dl}.
\end{equation}
Hence:
\begin{equation}\label{finalmente}
    \mathcal E'=\Gamma \mathcal E.
\end{equation}
   We have thus shown that the phenomenon of electromagnetic induction, involving electric and magnetic fields, must be treated relativistically, as claimed by Einstein \cite{ein05r}. Provided that the relative velocity is such that $V\ll c$, we can assume $ \Gamma \simeq 1 $, and
 the predicted value of $ \Gamma $ differs from $1$ by an
 experimentally not detectable amount.


\vskip5mm
\begin{thebibliography}{}
\bibitem{faraday1} M. Faraday,  {\em Experimental Researches in Electricity} vol. \href{https://archive.org/details/b21497916/page/n3/mode/2up} {I}, pp. 1-16 (Taylor, London, 1839).
\bibitem{lenz} E . Lenz,  ``\"{U}ber die Bestimmung der Richtung
der durch elektrodynamische Vertheilung erregten galvanischen
Str\"{o}me,'' {Ann. Phys. Chem.}
\href{https://gallica.bnf.fr/ark:/12148/bpt6k151161/f499.item.r=lenz.langEN}{{\bf
31 } 483-494} (1834).
\bibitem{treatise2}J. C. Maxwell,   {\em A Treatise on Electricity and
Magnetism}
\href{https://archive.org/details/electricityndmag02maxwrich/page/n5/mode/2up}{vol.
II} sec. ed. (Clarendon Press, Oxford, 1881).

\bibitem{feyn2}R. Feynman, R. Leighton  and M. Sands,    {\em The Feynman lectures on Physics} \href{http://www.feynmanlectures.caltech.edu/II_toc.html}{vol. II} pp. 17.1-17.3 (Addison~-~Wesley, Reading, 1963).
\bibitem{max_598}  A. D. Yaghjian,  ``Maxwell's derivation of the Lorentz force from
Faraday's law,'' Progress In Electromagnetics Research M, \href{https://www.jpier.org/pierm/pier.php?paper=20040202} {{\bf 93}, 35-42} (2020).
\bibitem{bouasse} H. Bouasse,  {\em Cours de Magn\'etisme et d'\'Electricit\'e - Premi\`ere Partie - \'Etude du Champ Magn\'etique} (Delagrave, Paris, 1914), online \href{https://archive.org/details/coursdemagnetism00p1unse/mode/2up}{here}.
   \bibitem{treatise1} J. C. Maxwell,  {\em A Treatise on Electricity and
Magnetism}
\href{https://archive.org/details/electricityndmag01maxwrich/page/n5/mode/2up}{vol. I} 2nd. ed.
(Clarendon Press, Oxford, 1881).
 \bibitem{symbol}Indeed, this symbol replaces, for technical reasons, the symbol used by Maxwell in \cite[p. 72]{treatise1}.
    \bibitem{SM} See the script in supplementary material at \href{https://fisica.unipv.it/percorsi/pdf/SM.pdf}{Supplementary material}.
    \bibitem{oliverh}B. J. Hunt, ``Oliver Heaviside: A first-rate oddity'', Phys. Today \href{https://physicstoday.scitation.org/doi/pdf/10.1063/PT.3.1788}{{\bf 65} 48-54} (2012).
\bibitem{epl} G. Giuliani,    ``A general law for electromagnetic
induction,'' {EPL}
\href{http://iopscience.iop.org/article/10.1209/0295-5075/81/60002?fromSearchPage=true}{{\bf
81} 60002} (2008).
\bibitem{names}The name ``electromotive force'' is misleading, because, as we know, the
electromotive force is not a force: it has the dimensions of an
electric potential, and it is measured in volts. Since a more
suitable name has not been invented, we shall keep on using the
same name. It is worth noting that, today's use of some historical names appears to
have no justification. For instance,
we find that the magnetic field  $\vec B $,
recovering a nineteenth  century's denotation, is called ``magnetic
induction vector''; and the field $ \vec H $, whose sources are the
current densities $ \vec J $,  is called  ``magnetic field''. Then one
has to stress that what appears in the expression of the Lorentz
force is the magnetic induction vector and not the magnetic field.
Let us also mention the conceptual  confusion created by the habit of
recalling the contributions of different researchers in the name
of a formula, a habit that often badly distorts history.
 For instance, as for the ``flux rule'', we have encountered the denomination ``Faraday - Neumann - Lenz law'', or variants at will.
The presence of Lenz is justified by the sign (-) that appears
in the formula, but Faraday and Neumann  have nothing to do with the ``flux rule''.
\bibitem{sf} J. C. Slater  and N. H. Frank,  {\em Electromagnetism} (McGraw - Hill, New York, 1947).
\bibitem{causality} As it is well known, an equation is local if it
relates physical quantities at the same point  at the same
instant, or if it relates physical quantities in two distinct points
at two successive instants $t_1, t_2$, provided that the
distance $d$ between the two points satisfies the equation:
$d\le c(t_2-t_1)$. This locality condition is necessary but not
sufficient for interpreting an equation causally. For
instance, let us consider the law $\vec F = d\vec p/dt$. As the
momentum varies over time, we are inclined to interpret this
equation by saying that the force $ \vec F $ `causes' the
momentum variation. However, there are
situations where the change in momentum `causes'  a force.
Consider, for example, a completely absorbing surface $ S $ hit
perpendicularly by a monochromatic beam of light
 directed along the negative direction of the $x$ axis. In this
case, if $N$ is
the number of photons absorbed per unit time, the surface momentum $P_x$ obeys:
$ {dP_x}/{dt}=-N
{h\nu}/{c}  =F_x$, and  we can say that the variation of the photons'
momentum has produced a radiative force $ F_x $ on the
surface $ S $. A similar situation is found in the kinetic theory of
gases: the pressure (force per unit area) on the walls is due to the
exchange of momentum with the particles.
\bibitem{ein05r}
A. Einstein,  ``Zur Electrodynamik bewegter K\"{o}rper,'' {\em Ann. Phys.} \href{https://onlinelibrary.wiley.com/doi/abs/10.1002/andp.19053221004} {{\bf 17} 891 - 921} (1905), English. trans. in {\em The collected papers of Albert Einstein}   vol 2 (Princeton,
NJ: Princeton University Press)
\href{http://einsteinpapers.press.princeton.edu/vol2-trans/154}{140
- 171} p. 140.

\bibitem{andy} A. Zangwill,  {\em Modern Electrodynamics}
(Cambridge University Press, Cambridge, 2013).
\bibitem{blondel} A. Blondel,   ``Sur l'\'enonc\'e le plus general des lois de l'induction,'' {Compt. Rend. Ac. Sc.}
\href{https://gallica.bnf.fr/ark:/12148/bpt6k3112q/f672.image}{\textbf{159} 674 - 679} (1914).
\bibitem{ggvp} G. Giuliani, ``Vector potential, electromagnetic
    induction and `physical meaning','' {Eur. J. Phys.}
    \href{http://dx.doi.org/10.1088/0143-0807/31/4/017}{{\bf 31} 871 -
    880} (2010).
    \bibitem{griff}D. J. Griffiths, {Introduction to Electrodynamics} 4th ed. (Pearson, Boston, 2013).
    \bibitem{rotore} The above discussion reminds us of another question.
Let us consider
Maxwell's  equation:
\begin{equation}\label{eqmax}
\nabla \times {\vec E}=-\frac{\partial \vec B}{\partial t}.
\end{equation}
It is easy to find  statements according to which this equation
shows that a time-varying magnetic field {\em causes} an electric
field (and symmetrical statements for the equation of the {\em curl}
of the magnetic field). See, for instance \cite{hill}, where
statements of this kind are considered to be conceptual
misconceptions that one should avoid in teaching.  Equation (\ref{eqmax}) states only a {\em relation} between the fields
  as the charges produce them: see also \cite{jef}. \label{puntiform}This is well illustrated by the equations that yield the fields
produced by a point charge in arbitrary motion. The electric and magnetic fields are given by equations that
depend on the charge's velocity and acceleration. These equations are independently deduced one from
the other (see, for instance, \cite[pp. 870-879]{andy}. Only ex-post, do we find that the two fields are {\em related} by the equation:
\begin{equation}\label{pointmag}
\vec B = \frac{1}{c}(\hat r^*_{21}\times \vec E),
\end{equation}
      where $\hat r^*_{21}$ is the unit vector pointing from the retarded position of the point charge towards the point at which the field is calculated.
In the same spirit, let us emphasize that the presentation of Maxwell's equations in integral form deals a mortal blow to the intrinsic local nature of Maxwell's theory, and it transforms the production of electromagnetic waves into a profound mystery. These equations connect what happens on a closed line at time  $t$ to what happens simultaneously on an arbitrary surface having the line as a contour. This habit is widespread in textbooks for high school in Italy but not -- for instance -- in textbooks published in the United States.

    \bibitem{hill} S. E. Hill,    ``Rephrasing Faraday's Law,'' {Phys. Teach.} \href{http://link.aip.org/link/doi/10.1119/1.3479724?ver=pdfcov}{{\bf 48} 410 - 412} (2010).
\bibitem{jef} O. D. Jefimenko,  ``Presenting electromagnetic theory in accordance
with the principle of causality,'' {Eur. J. Phys.} \href{https://iopscience.iop.org/article/10.1088/0143-0807/25/2/015}{{\bf 25} 287 - 296} (2004).
\bibitem{heavi}O. Heaviside, {\em Electromagnetic Theory}, \href{https://archive.org/details/electromagnetict0001heav/mode/2up}{vol. I}, (`The Electrician' Printing and Publishing Company, London, 1893).
    \bibitem{ew}H. Hertz, {\em Electric waves}, (McMillan and CO, London, 1893), p.196.
\bibitem{kuhn} T. S. Kuhn, {\em The Structure of Scientific Revolutions} 2nd ed. (University of Chicago Press,  Chicago, 1970).
    \bibitem{riecke}E. Riecke,  {\em Lehrbuch der Experimental Physik - Zweiter Band, Magnetismus, Elektrizit\"{a}t. W\"{a}rme} (Verlag
von Veit \& Co.,  Leipzig, 1896) pp. 196 - 226; 254 - 268, online \href{https://archive.org/details/lehrbuchderexpe00riecgoog}{here}.
\bibitem{landau_ecm} L. L. Landau and E. M. Lifshitz,   {\em Electrodynamics of
Continuous Media} \href{https://archive.org/details/ElectrodynamicsOfContinuousMedia}{(Pergamon Press, Oxford, 1960)} pp. 205-209.
\bibitem{jackson}J. D. Jackson,   {\em Classical Electrodynamics} 3rd ed. (John Wiley \& Sons, New York, 1999) pp. 208-211.
\bibitem{rousseaux}   G. Rousseaux,  R. Kofman  and  O.
    Minazzoli,   ``The Maxwell - Lodge effect: significance of
    electromagnetic potentials in the classical theory,'' {Eur. Phys.
    J.} D \href{http://dx.doi.org/10.1140/epjd/e2008-00142-y} {{\bf 49}
    249 - 256} (2008).
    \bibitem{pp} W. K. H. Panofsky  and M. Phillips,   {\em Classical Electricity andf Magnetism} 2nd ed. (Addison - Wesley, Reading, 1955).
         \bibitem{purmor} E. M. Purcell  and D. J. Morin,   {\em Electricity and Magnetism} 3rd ed. (Cambridge University Press, Cambridge, 2013).
       \bibitem{high_school} The applicability of this proposal to high schools depends on the mathematical background of the students.  It is likely applicable in Italy's scientific Lyceums, where mathematical knowledge is sufficient to treat electromagnetic induction as here suggested.
                 \bibitem{labexp} G. Giuliani, ``L'induzione elettromagnetica: un percorso didattico'', Gior. Fis. {\bf 49}, 291-304, (2008).
                 \bibitem{fleck} L. Fleck, {\em Genesis and Development of a Scientific Fact}, (The University of Chicago Press, Chicago, 1979).
\end{thebibliography}
\end{document}